\documentclass{ws-p9-75x6-50}
\usepackage{graphicx,amssymb,amsthm}
\newcommand{\ua}{\uparrow}
\newcommand{\da}{\downarrow}

\begin{document}

\title{Evolution of fermionic systems as 
\\ an expectation over Poisson processes}

\author{M. Beccaria}
\address{Dipartimento di Fisica, Universit\`a di Lecce, Via Arnesano,
73100 Lecce, Italy}
\author{C. Presilla}
\address{Dipartimento di Fisica, Universit\`a di Roma ``La Sapienza'',
Piazzale A. Moro 2, 00185 Roma, Italy}
\author{G. F. De Angelis}
\address{Dipartimento di Fisica, Universit\`a di Lecce, Via Arnesano,
73100 Lecce, Italy}
\author{G. Jona-Lasinio}
\address{Dipartimento di Fisica, Universit\`a di Roma ``La Sapienza'',
Piazzale A. Moro 2, 00185 Roma, Italy}

\maketitle

\abstracts{
We derive an exact probabilistic representation for the evolution of a
Hubbard model with site- and spin-dependent hopping 
coefficients and site-dependent interactions in terms of an associated 
stochastic dynamics of a collection of Poisson processes.
}

In a recent paper\cite{BPDAJL99} we provided an exact probabilistic 
expression for the real time or the imaginary time evolution 
of a Fermi system, in particular a Hubbard model, in terms of an 
associated stochastic dynamics of a collection of Poisson processes.
Here, we generalize the result\cite{BPDAJL99} to the case of 
a Hubbard model with site- and spin-dependent hopping coefficients 
and site-dependent interactions.
This situation is of interest if disorder is present\cite{SSH94}.

Let us consider the Hubbard Hamiltonian
\begin{eqnarray}
\label{Hubbard}
H &=& - \sum_{i=1}^{|\Lambda|} \sum_{j=i}^{|\Lambda|} 
\sum_{\sigma=\ua\da} \eta_{ij\sigma}
(c^\dag_{i\sigma} c^{}_{j\sigma} + c^\dag_{j\sigma} c^{}_{i\sigma}) 
+\sum_{i=1}^{|\Lambda|} \gamma_i~ 
c^\dag_{i\ua} c^{}_{i\ua}~c^\dag_{i\da} c^{}_{i\da},
\end{eqnarray}
where $\Lambda\subset Z^d$ is a finite $d$-dimensional lattice
with cardinality $|\Lambda|$,
$\{1, \dots,|\Lambda|\}$ some total ordering of the lattice points,
and $c_{i \sigma}$  the usual anticommuting destruction operators
at site $i$ and spin index $\sigma$.
Note that the Hamiltonian (\ref{Hubbard}) also allows for 
spin-dependent site energies $-2\eta_{ii\sigma}$.
We are interested in evaluating the matrix elements
$\langle {\bf n}' |e^{- iHt} | {\bf n} \rangle$
where ${\bf n}= (n_{1 \ua},n_{1 \da}, \ldots, n_{|\Lambda| \ua},
n_{|\Lambda| \da})$ are the occupation numbers taking the values 0 or 1.
Since the total number of fermions per spin component is a conserved
quantity, we consider only configurations ${\bf n}$ and 
${\bf n}'$ such that 
$\sum_{i=1}^{|\Lambda|} n'_{i\sigma} = \sum_{i=1}^{|\Lambda|} n_{i\sigma}$
for $\sigma = \ua \da$.
In the following we shall use the modulus 2 addition
$n \oplus n'= (n+n') \bmod 2$.

Let $\Gamma_\sigma =\{(i,j), 1\le i< j\le |\Lambda|: 
\eta_{ij\sigma}\neq 0\}$ and $|\Gamma_\sigma|$ its cardinality.
By introducing
\begin{eqnarray}
\label{lambda}
\lambda_{ij \sigma}({\bf n}) &\equiv& 
\langle {\bf n} \oplus {\bf 1}_{i\sigma} \oplus {\bf 1}_{j\sigma}|
c^\dag_{i\sigma} c^{}_{j\sigma} + c^\dag_{j\sigma} c^{}_{i\sigma}
|{\bf n}\rangle 
\nonumber \\ &=&
(-1)^{n_{i\sigma} + \cdots + n_{j-1 \sigma}} 
\left[ n_{j \sigma} (n_{i \sigma} \oplus 1) -
n_{i \sigma} (n_{j \sigma} \oplus 1) \right],  
\end{eqnarray}
where ${\bf 1}_{i\sigma}=(0,\ldots,0,1_{i\sigma},0,\ldots,0)$, and 
\begin{eqnarray}
V({\bf n}) &\equiv& 
\langle {\bf n}|H|{\bf n}\rangle =
\sum_{i=1}^{|\Lambda|} \gamma_i n_{i \ua} n_{i \da}
-\sum_{i=1}^{|\Lambda|} \sum_{\sigma=\ua\da} 
2\eta_{ii\sigma} n_{i\sigma}, 
\label{potential}
\end{eqnarray}
the following representation holds
\begin{eqnarray}
\label{TheFormulaa}
\langle {\bf n}'|e^{-iHt} | {\bf n}\rangle &=&  {\bf E} \left(
\delta_{ {\bf n}' , {\bf n} \oplus {\bf N}^t } 
{\cal M}^t \right) 
\end{eqnarray}
\begin{eqnarray}
\label{TheFormulab}
{\cal M}^t &=& \exp \biggl\{ 
\sum_{\sigma=\ua\da} \sum_{(i,j)\in\Gamma_\sigma} 
\int_{[0,t)} \!\!\!\!\!\!
\log \left[ i\eta_{ij\sigma} \rho_{ij\sigma}^{-1} 
\lambda_{ij \sigma} ({\bf n} \oplus {\bf N}^s) \right] dN^s_{ij\sigma} 
\nonumber \\ && 
- i\int_0^t V({\bf n} \oplus {\bf N}^{s}) ds 
+ \sum_{\sigma=\ua\da} \sum_{(i,j)\in\Gamma_\sigma} \rho_{ij\sigma} t 
\biggr\}.
\end{eqnarray}
Here, $\{N_{ij\sigma}^t\}$, $(i,j) \in \Gamma_\sigma$ and 
$\sigma=\ua\da$, is a family of 
$|\Gamma_\ua|+|\Gamma_\da|$ independent Poisson processes with parameters 
$\rho_{ij\sigma}$ and 
${\bf N}^t= (N_{1 \ua}^t,N_{1 \da}^t, \ldots, N_{|\Lambda| \ua}^t,
N_{|\Lambda| \da}^t)$ are $2|\Lambda|$ stochastic processes defined as
\begin{equation}
N^t_{k \sigma} = \sum_{
(i,j) \in \Gamma_\sigma:~i=k~{\rm or}~j=k }
N^t_{ij \sigma}.
\end{equation}
We remind that a Poisson process $N^t$ with parameter $\rho$ is a jump 
process characterized by the probabilities
$P\left( N^{t+s} - N^t=k \right) = (\rho s)^k e^{-\rho s} /k!$.
Its trajectories are piecewise-constant increasing integer-valued 
functions continuous from the left.
The stochastic integral $\int dN^t$ is just an ordinary 
Stieltjes integral
\begin{eqnarray}
\int_{[0,t)} f(s,N^s) dN^s = \sum_{k:~ s_k <t} f(s_k,N^{s_k}),
\nonumber
\end{eqnarray}
where $s_k$ are random jump times having probability density
$p(s)=\rho e^{- \rho s}$.
Finally, the symbol ${\bf E}( \ldots )$ is the expectation of the 
stochastic functional within braces.

The matrix elements $\langle {\bf n}'|e^{-iHt} | {\bf n}\rangle$
obey the ODE system
\begin{eqnarray}
{d \over dt}
\langle {\bf n}'|e^{-iHt} | {\bf n}\rangle =  
- i\sum_{{\bf n}''} \langle {\bf n}'| H | {\bf n}''\rangle
\langle {\bf n}''| e^{-iHt} | {\bf n}\rangle
\label{ODE}
\end{eqnarray}
with initial condition 
$\left. \langle {\bf n}'|e^{-iHt} | {\bf n}\rangle \right|_{t=0}=
\delta_{{\bf n}' {\bf n}}$.
One may check that (\ref{TheFormulaa}-\ref{TheFormulab}) is indeed
solution of (\ref{ODE}) by applying the rules of stochastic 
differentiation. 
We have
\begin{eqnarray}
&{\bf E}& \left( 
\delta_{ {\bf n}' , {\bf n} \oplus {\bf N}^{t+dt}}
{\cal M}^{t+dt} \right) 
\nonumber \\ &&= 
\sum_{{\bf n}''} {\bf E} \biggl( 
\prod_{\sigma=\ua\da} \prod_{(i,j)\in\Gamma_\sigma}
\delta_{ {\bf n}' , {\bf n}'' \oplus d {\bf N}^t}
e^{ \int_{[t,t+dt)}
\log \left[ i\eta_{ij\sigma} \rho_{ij\sigma}^{-1} \lambda_{ij\sigma} 
({\bf n} \oplus {\bf N}^s) \right] dN^s_{ij\sigma} }  
\nonumber \\ &&~~\times
e^{- iV({\bf n} \oplus {\bf N}^{t}) dt 
+ \sum_{\sigma=\ua\da} \sum_{(i,j)\in\Gamma_\sigma} \rho_{ij\sigma} dt} 
~\delta_{ {\bf n}'' , {\bf n} \oplus {\bf N}^t} {\cal M}^t \biggr)
\label{dM}
\end{eqnarray}
For the Markov property, the expectation of the factors 
containing the stochastic integrals in the interval $[t,t+dt]$ can
be taken separately. 
By expanding each one of them over all possible numbers of
jumps of the Poisson processes as
\begin{eqnarray}
&{\bf E} &\left( \delta_{ {\bf n}' , {\bf n}'' \oplus d {\bf N}^t}
~e^{\int_{[t,t+dt)} 
\log \left[ i\eta_{ij\sigma} \rho_{ij\sigma}^{-1} \lambda_{ij\sigma} 
({\bf n} \oplus {\bf N}^s) \right] dN^s_{ij\sigma} } \right) 
\nonumber \\ &&=
\delta_{ {\bf n}' , {\bf n}'' } ~e^0~e^{-\rho_{ij\sigma} dt} 
+ \delta_{ {\bf n}' , {\bf n}'' \oplus {\bf 1}_{i\sigma}
\oplus {\bf 1}_{j\sigma} }
~e^{\log \left[ i\eta_{ij\sigma} \rho_{ij\sigma}^{-1} \lambda_{ij\sigma} 
({\bf n} \oplus {\bf N}^t) \right] } 
~e^{-\rho_{ij\sigma} dt} \rho_{ij\sigma} dt + \ldots
\nonumber \\ &&=
\delta_{ {\bf n}' , {\bf n}'' } + 
\left[ \delta_{ {\bf n}' , {\bf n}'' \oplus {\bf 1}_{i\sigma}
\oplus {\bf 1}_{j\sigma} }
i\eta_{ij\sigma} \rho_{ij\sigma}^{-1} 
\lambda_{ij\sigma}({\bf n} \oplus {\bf N}^t) \right.
\left. - \delta_{ {\bf n}' , {\bf n}'' } \right] \rho_{ij\sigma} dt 
+ {\cal O}\left( dt^2 \right),
\nonumber
\end{eqnarray}
to order $dt$ we obtain
\begin{eqnarray}
{\bf E} \left( 
\delta_{ {\bf n}' , {\bf n} \oplus {\bf N}^{t+dt}}
{\cal M}^{t+dt} \right) 
&=&
\sum_{{\bf n}''} \biggl[ \delta_{ {\bf n}' , {\bf n}''} 
+\sum_{\sigma=\ua\da} \sum_{(i,j)\in\Gamma_\sigma}
\delta_{ {\bf n}' , {\bf n}'' \oplus {\bf 1}_{i\sigma}
\oplus {\bf 1}_{j\sigma} }
i\eta_{ij\sigma} \lambda_{ij\sigma} ({\bf n}'') dt
\nonumber \\ &&- 
\delta_{ {\bf n}' , {\bf n}''} iV({\bf n}'') dt \biggr] 
{\bf E} \left(
\delta_{ {\bf n}'' , {\bf n} \oplus {\bf N}^t}
 {\cal M}^t \right).
\end{eqnarray}  
Finally, we rewrite this relation as
\begin{eqnarray}
d {\bf E} \left( 
\delta_{ {\bf n}' , {\bf n} \oplus {\bf N}^t}
{\cal M}^t \right) 
&=& 
{\bf E} \left( 
\delta_{ {\bf n}' , {\bf n} \oplus {\bf N}^{t+dt}}
{\cal M}^{t+dt} \right) 
- {\bf E} \left( 
\delta_{ {\bf n}' , {\bf n} \oplus {\bf N}^t}
{\cal M}^t \right) 
\nonumber \\ &=&
- i\sum_{{\bf n}''} 
\langle {\bf n}'| H | {\bf n}'' \rangle
{\bf E} \left( \delta_{ {\bf n}'' , {\bf n} \oplus {\bf N}^t } 
 {\cal M}^t  \right) dt. 
\label{final}
\end{eqnarray}
It is clear that the fermionic nature of $H$ plays no special role 
in the above derivation which holds for any system described by a
finite Hamiltonian matrix.

In order to construct an efficient algorithm for evaluating 
(\ref{TheFormulaa}-\ref{TheFormulab}), 
we start by observing that the functions 
$\lambda_{ij\sigma}({\bf n} \oplus {\bf N}^s)$ vanish 
when the occupation numbers $n_{i\sigma} \oplus N_{i\sigma}^s$ and 
$n_{j\sigma} \oplus N_{j\sigma}^s$ are equal.
We say that for a given value of $\sigma$ the link $ij$ is active
at time $s$ if $\lambda_{ij\sigma}({\bf n} \oplus {\bf N}^s)\neq 0$.
We shall see in a moment that only active links are relevant. 
Let us consider how the stochastic integral in (\ref{TheFormulab}) builds
up along a trajectory defined by considering the time ordered 
succession of jumps in the family $\{ N^t_{ij\sigma} \}$.
The contribution to the stochastic integral in the 
exponent of (\ref{TheFormulab}) at the time of the first jump, 
for definiteness suppose that the link $i_1j_1$ with spin component
$\sigma_1$ jumps first at time $s_1$, is
\begin{eqnarray}
\log \left[ i\eta_{i_1j_1\sigma_1} \rho_{i_1j_1\sigma_1}^{-1} 
\lambda_{i_1j_1\sigma_1}({\bf n} \oplus {\bf N}^{s_1}) \right] 
~\theta(t-s_1),
\nonumber
\end{eqnarray}
where ${\bf N}^{s_1}={\bf 0}$ due the assumed left continuity.
Therefore, we obtain a real finite or $-\infty$ contribution to 
the stochastic integral depending if the link $i_1j_1\sigma_1$ was 
active or not at time 0. 
If $s_1 \geq t$ we have no contribution to the stochastic 
integral from this trajectory. 
If $s_1 < t$ a second jump of a link, suppose $i_2j_2$ with spin
component $\sigma_2$, can take place at time $s_2>s_1$.
The analysis can be repeated by considering an arbitrary number of jumps.
Of course, when the real part of the stochastic integral is $- \infty$, 
i.e. when some $\lambda=0$, there is no contribution to the expectation. 
The other integral in (\ref{TheFormulab}) is an ordinary integral of
a piecewise constant bounded function. 

From the above remarks it is clear that the only trajectories to be 
considered are those associated to the jumps of active links.
We start by determining the set of active links 
$A_\sigma^1 \subset \Gamma_\sigma$, $\sigma=\ua\da$, 
in the initial configuration ${\bf n}$ assigned at time $0$.
This is done by 
inspecting the occupation numbers of the sites according to the 
rule that the link $ij$ is active for the spin component $\sigma$ if 
$n_{i\sigma} + n_{j \sigma} =1$. 
Then, we extract the jump time $s_1$ according to the probability
density 
\begin{equation}
p(s)= 
\sum_{\sigma=\ua\da} \sum_{(i,j)\in A_\sigma^1} 
\rho_{ij\sigma} 
~e^{-\sum_{\sigma=\ua\da} \sum_{(i,j)\in A_\sigma^1} 
\rho_{ij\sigma}s}. 
\end{equation}
Finally, we choose the jumping link in the set $A_\sigma^1$
with probability 
\begin{equation}
P_{ij\sigma} =
\frac{ \rho_{ij\sigma} }
{\sum_{\sigma=\ua\da} \sum_{(i,j)\in A_\sigma^1} 
\rho_{ij\sigma}  }.
\end{equation}
The contribution to ${\cal M}^t$ at the time of the first jump 
is therefore, up to the last factor which appears in (\ref{TheFormulab}),
\begin{eqnarray}
&& i\eta_{i_1j_1 \sigma_1} \rho_{i_1j_1 \sigma_1}^{-1} 
\lambda_{i_1j_1 \sigma_1} ({\bf n} \oplus 
{\bf N}^{s_1})
e^{ -iV({\bf n} \oplus {\bf N}^{s_1}) s_1 }
e^{-\sum_{\sigma=\ua\da} \sum_{(i,j)\in \Gamma_\sigma \setminus
A_\sigma^1} \rho_{ij\sigma} s_1} \theta(t-s_1)
\nonumber \\ && 
+ e^{ -iV({\bf n} \oplus {\bf N}^{t}) t }
e^{-\sum_{\sigma=\ua\da} \sum_{(i,j)\in \Gamma_\sigma \setminus 
A_\sigma^1} \rho_{ij\sigma} t} ~\theta(s_1-t),  
\nonumber
\end{eqnarray}
where $\exp \left( -\sum_{\sigma=\ua\da} 
\sum_{(i,j)\in \Gamma_\sigma \setminus A_\sigma^1} 
\rho_{ij\sigma} s \right)$
is the probability that the non active links do not jump in the 
time interval $s$.
The contribution of a given trajectory is obtained by multiplying the
factors corresponding to the different jumps until the last jump 
takes place later than $t$.
For a given trajectory we thus have 
\begin{eqnarray}
{\cal M}^t &=& \prod_{k \geq 1}
\Bigl [
i\eta_{i_kj_k \sigma_k} \rho_{i_kj_k \sigma_k}^{-1} 
\lambda_{i_kj_k \sigma_k} ({\bf n} \oplus {\bf N}^{s_k})
e^{[\zeta_k-iV({\bf n} \oplus {\bf N}^{s_k})] (s_k-s_{k-1}) }
~\theta(t-s_k) 
\nonumber \\ &&+
e^{[\zeta_k-iV({\bf n} \oplus {\bf N}^{t})] (t-s_{k-1}) }
~\theta(s_k-t) \Bigr].
\label{Mt}
\end{eqnarray}
Here, $\zeta_k=
\sum_{\sigma=\ua\da} \sum_{(i,j)\in A_\sigma^k} \rho_{ij\sigma}$
where $A_\sigma^k$ is the set of active links with spin $\sigma$
in the interval $(s_{k-1}, s_k]$ and $s_0=0$.
Note that the last exponentially increasing factor in 
(\ref{TheFormulab}) cancels out in the final expression of ${\cal M}^t$.
The analogous expression of ${\cal M}^t$ for imaginary times is simply 
obtained by replacing $\eta \to -i \eta$ and $\gamma \to -i \gamma$.

In principle, the algorithms parametrized by different 
$\rho_{ij\sigma}$ are all equivalent
as (\ref{TheFormulaa}-\ref{TheFormulab}) holds for any choice of
the Poisson rates.
However, since we estimate numerically the expectation values 
with a finite number of trajectories, this may introduce a systematic 
error.
It can be shown that the best performance is obtained for the 
natural choice $\rho_{ij\sigma} \sim \eta_{ij\sigma}$ 
independently of the interaction strength $\gamma$.

\end{document}